\documentclass{article}

\usepackage{euscript}

\usepackage{latexsym}
\usepackage{amsmath, amssymb,graphics}
\usepackage[all]{xy}

\usepackage{fullpage}

\newtheorem{theorem}{Theorem}

\newtheorem{lemma}[theorem]{Lemma}

\newtheorem{proposition}[theorem]{Proposition}

\renewcommand{\(}{\begin{equation*}}
\renewcommand{\)}{\end{equation*}}
\newcommand{\bea}{\begin{eqnarray*}}
\newcommand{\eea}{\end{eqnarray*}}
\newcommand{\R}{{\mathbb R}}
\newcommand{\C}{{\mathbb C}}
\newcommand{\Z}{{\mathbb Z}}

\newcommand{\D}{{\mathbb D}}

\newcommand{\cC}{\ensuremath{\mathcal C}}

\newcommand{\cH}{\ensuremath{\mathcal H}}

\newcommand{\bo}{\raise-1mm\hbox{\Large$\Box$}}              
\newcommand{\cS}{{\mathcal S}}

\def\D{\ensuremath{{\cal D}}}

\newcommand{\beq}{\begin{equation}}
\newcommand{\eeq}{\end{equation}}

\numberwithin{equation}{section}

\renewcommand{\(}{\begin{equation}}
\renewcommand{\)}{\end{equation}}

\newcommand{\CC}{{\mathbb C}}

\newcommand{\CP}{\CC \text{P}}

\def\R{{\mathbb R}}
\def\Z{{\mathbb Z}}

\def\C{{\mathbb C}}

\def\1{{\bf 1}}

\def\<{\langle}
\def\>{\rangle}

\numberwithin{equation}{section}

\renewcommand{\(}{\begin{equation}}
\renewcommand{\)}{\end{equation}}

\begin{document}

\begin{titlepage}


\vspace{2em}
\def\thefootnote{\fnsymbol{footnote}}

\begin{center}
{\Large\bf 
Corners in M-theory}
\end{center}
\vspace{1em}

\begin{center}
\large Hisham Sati 
\footnote{e-mail: {\tt
hsati@math.umd.edu}}
\end{center}

\begin{center}
Department of Mathematics\\
University of Maryland\\
College Park, MD 20742 
\end{center}

\vspace{0em}
\begin{abstract}
\noindent

M-theory can be defined on closed manifolds as well as on 
manifolds with boundary. As an extension,
we show that manifolds with corners appear naturally in M-theory. 
We illustrate this with four situations: The lift to bounding twelve dimensions 
of M-theory on Anti de Sitter spaces, ten-dimensional 
heterotic string theory in relation to twelve dimensions, and 
the two M-branes 
within M-theory in the presence of
a boundary. The M2-brane is taken with (or as) a boundary and the 
  worldvolume of the M5-brane is viewed
as a tubular neighborhood. We then concentrate on (variant) of the heterotic 
theory as a corner and explore analytical and geometric consequences. 
In particular, we formulate and study the phase of the partition function 
in this setting 
and identify the corrections due to the corner(s). 
The analysis involves considering M-theory on 
disconnected manifolds, and makes use of the extension of the 
Atiyah-Patodi-Singer index theorem to manifolds with corners
and 
the $b$-calculus of Melrose.

\end{abstract}

\end{titlepage}

\tableofcontents

\section{Introduction}

The boundary of a manifold $X$ cannot itself be a boundary, but 
there are situations in physics (and mathematics) which 
demand that sense be made out of some variation on the 
notion of boundaries of boundaries. 
 For example, we know that heterotic
string theory should be a boundary of M-theory \cite{HW1} \cite{HW2}.
The latter in turn is best described globally using a bounding twelve-dimensional 
theory \cite{Flux} \cite{DMW}. A naive boundary of a boundary does 
not exist as e.g. the boundary operator in homology is nilpotent. However, 
this has a natural setting within manifolds with corners. 
With this we then can view the heterotic theory  as codimension-two corner of the 
twelve-dimensional theory.

\vspace{3mm}
We recall the basics of manifolds with corners in section \ref{sec base}.
Closed manifolds and manifolds with boundary are special
cases of manifolds with corners. Also products of two manifolds with 
boundary form an interesting class of examples within
manifolds with corners of codimenison-two
We consider situations in M-theory, in addition to the heterotic boundary, 
where manifolds with corners are
needed in order to describe the physical system. This is discussed
 in section \ref{sec m} and includes the following.

\paragraph{AdS/CFT correspondence.}
Eleven-dimensional supergravity admits solutions with an Anti-de Sitter space
as a factor, most famously  
${\rm AdS}_4 \times S^7$  \cite{FR} and 
${\rm AdS}_7 \times S^4$
\cite{PNT}. These spaces appear as the near horizon limits of the M2-brane and 
the M5-brane, respectively,  via the AdS/CFT correspondence which relates
quantum supergavity on the AdS factor to the conformal field theory on the 
boundary $\partial {\rm AdS}$ \cite{Ma}. In section \ref{sec ads}
we show that the lift of these solutions to twelve dimensions
leads naturally to a manifold with corners as the product of two manifolds with
boundaries.

\paragraph{The M5-brane.}
The M5-brane with worldvolume $W^6$  can be 
described in terms of tubular neighborhoods in the target
11-dimensional manifold $Y^{11}$. When $Y^{11}$ has no boundary the resulting 
manifold arising from the sphere bundle of the normal bundle to the embedding 
$W^6 \hookrightarrow Y^{11}$ has a boundary. Now M-theory itself on a manifold with 
boundary certainly makes sense \cite{HW2} \cite{DFM} \cite{DMW-boundary} and so 
we ask what happens to the description of the M5-brane in that case. 
While the main theme in 
\cite{DFM}  \cite{DMW-boundary} was for  when $Y^{11}$ has a boundary, 
  the description of the M5-brane was given only for the case 
  when $Y^{11}$ is closed
  (in \cite{DFM} the M5-brane was related to M-theory with boundary only via
  anomalies involving torsion).
In section \ref{sec M5}
we show how the resulting manifold will be a manifold with corners of codimension-two.
This uses the general 
result of \cite{Ja} that the removal of a tubular neighborhood 
of any submanifold creates a manifold of one codimension higher.
Therefore, including an M5-brane
on an eleven-dimensional manifold with a boundary leads naturally to 
a manifold with corners.

\paragraph{The M2-brane.}
The M2-brane in M-theory can have boundaries on the M5-brane \cite{Str}.
When M-theory is considered on a manifold with boundary, then the 
two-dimensional boundaries of the M2-brane can end on the ten-dimensional 
boundary of M-theory, in the Horava-Witten set-up \cite{HW2}.
Therefore, we take the M2-brane within M-theory with boundary, that is on 
a twelve manifold which is a product of a three-manifold with boundary with 
an eight manifold. Then we wrap the M2-brane on the former; 
the ten-manifold
which is the product of the membrane boundary with the eight-manifold
will be a corner for the twelve-manifold.  In fact, this will be essentially the 
heterotic corner. We describe this in section \ref{sec M2}.

\paragraph{
Heterotic M-theory.}
Heterotic string theory is essentially a boundary of M-theory
\cite{HW1} \cite{HW2}.
The M-theory partition function on a Spin 
eleven-manifold with boundary was considered in \cite{DFM}
with an emphasis on eleven rather than on twelve-dimensions.  
For topological and global (e.g. index theory) purposes, M-theory in 
turn is considered as a boundary of the bounding twelve-dimensional theory on 
$Z^{12}$ \cite{Flux} \cite{DMW}. Hence,  
in the connection to heterotic string theory, the bounding twelve-dimensional 
theory requires having seemingly a `boundary of a boundary'. 
In section \ref{sec anal} we consider the effect of studying this from the point of view of the 
bounding twelve-dimensional theory.  We provide two formulations, one using Dirac operators 
(in section \ref{sec dir})
and another
involving the signature operators (in section \ref{sec sig}), making use of the emergence of the latter
in \cite{DMW-boundary}. 

\vspace{3mm}
As in \cite{F}, we consider the Horava-Witten theory
on a ten-dimensional Spin manifold 
$M^{10}$ from two points of view. First,  via the product with the interval $[0,1] \times M^{10}$
(the ``upstairs" formulation),
which for us nicely connects to manifolds with corners by taking a
 further Cartesian product with another interval. We do this for most of section \ref{sec anal}.
Second, via $S^1/\Z_2 \times M^{10}$ (the ``downstairs" formulation) 
with $\Z_2$ acting as an orientation-reversing involution.
We consider the eleven-manifold as a boundary of a twelve-manifold 
in the presence of an orientation-reversing involution and study the 
effect on the signature operator and the corresponding eta-invariants
in section \ref{sec rev}. This allows us to formulate the phase of the partition function.

\vspace{3mm}
We study analytical and geometric
 aspects of the theory
in this setting using mainly the constructions in \cite{Bu} \cite{HMM} and \cite{Mu}
and the survey \cite{Lo}.   
In particular we consider the global reduction to ten dimensions of the phase of the 
partition function, using $b$-eta-invariants within the $b$-calculus \cite{Mel}.
This allows for more general boundary conditions than those of Atiyah-Patodi-Singer
(APS) \cite{APS} used in \cite{DMW}.
The discussion requires considering M-theory on disconnected eleven-dimensional 
spaces. We also consider the case of multiple ten-dimensional (heterotic) components
in the setting of manifolds with corners. 

\vspace{3mm}
While this is mostly a physics paper, we have chosen to identify main (physical)
results and observations
by recording them as propositions and lemmas, mainly as a way of keeping track of the main 
statements.

\section{Manifolds with corners and their relevance in M-theory}
\label{sec cor}
We first recall in section \ref{sec base}
the basics of manifolds with corners and then we 
provide our applications to M-theory in section \ref{sec m}.

\subsection{Basic definitions 
and relevant tools}
\label{sec base}

We now give the basic definitions and some of the properties that we need in the applications
to M-theory, which we discuss starting in the following section.

\paragraph{The basic definitions.} 
 A {\it differentiable manifold with corners} 
is a topological space covered by charts which are locally 
open subsets of $\R^n_+=[0,\infty)^n$
\cite{Do} \cite{Ce}.
 Adding information about faces leads to 
{\it manifold with faces}.
 Imposing conditions on how the faces piece globally together 
leads to s restrictive class called $\langle n \rangle$-manifolds
\cite{Ja}. This is a manifold with faces together with an ordered $n$-tuple
$(\partial_0 X, \partial_1 X, \cdots, \partial_{n-1}X)$ of faces of $X$ which satisfy the 
following conditions:

\noindent $(1)$ The boundary is formed of $n$ disconnected components
$\partial X=\partial_0 X \cup \cdots \cup \partial_{n-1}X$;

\noindent $(2)$ The intersection  $\partial_i X \cap \partial_j X$ is 
a face of $\partial_iX$ and of $\partial_jX$ for all $i \neq j$.

\noindent The number $n$ is called the {\it codimension} of $X$.
We will be mainly interested in the case $n=2$.

\paragraph{Products and codimension.}
The product of an $\langle m\rangle$-manifold with an 
 $\langle n\rangle$-manifold 
  $\langle m+n \rangle$-manifold. 
 A $\langle 0 \rangle$-manifold is a manifold without a boundary while a
 $\langle 1 \rangle$-manifold is a manifold with boundary. 
So we can create many manifolds with boundary by multiplying manifolds of these 
two different types.
Furthermore, we can create $\langle n \rangle$-manifolds from 
products of $\langle 0 \rangle$-manifolds with $\langle n \rangle$-manifolds, 
$\langle 1\rangle$-manifolds with $\langle n-1\rangle$-manifolds, and so on.
In the main case of interest, which is 12-manifolds with corners of 
codimension-two, we can construct many such spaces by taking a product of a $k$-dimensional
 $\langle i \rangle$-manifold with a $(12-k)$-dimensional $\langle 2-i \rangle$-manifold
 with $i=0,1$. More explicitly, we can take the product of a closed manifold with 
 a manifold with corners as well as the product of two manifolds with boundary,
 the sum of whose dimensions is 12.

\vspace{3mm}
We consider two simple examples of manifolds with corners of codimension-two.

\paragraph{Example 1. The positive quadrant.}
Let $\overline{\R_+^2}$ denote the closed positive quadrant of $\R^2$, that is 
$\overline{\R_+^2}=\{ (x^1, x^2) \in \R^2 : x^1 \geq 0, x^2 \geq 0\}$.
The boundary of $\overline{\R_+^2}$ in $\R^2$ is the set of points at 
which one or both coordinates vanish. The points in $\overline{\R_+^2}$
at which both coordinates vanish are called its
{\it corner points}. 
The boundary of a smooth manifold with corners is in general not a smooth 
manifold with corners. For example, the boundary of $\overline{\R_+^2}$
is the union
$
\partial \overline{\R_+^2}= H_1 \cup H_2$,
where $H_i=\{ (x^1, x^2) \in \overline{\R_+^2}: x^i=0\}$, $i=1,2$, is a 
one-dimensional smooth manifold with boundary. 

\paragraph{Example 2. Lie groups with action of maximal torus.}
Let $G$ be $SU(2)$ or $SO(4)$, the Lie groups of rank 2, and let $T^2$ be the 
corresponding maximal torus. Then $T^2$ acts on the product $(\mathbb{D}^2)^2$ 
of 2 disks $\mathbb{D}^2$ by complex multiplication. The resulting  
associated fiber bundle $G \times_{T^2}  (\mathbb{D}^2)^2$ is a
$\langle 2 \rangle$-manifold. For $SU(2)$ this is five-dimensional,
while for $SO(4)$ this is eight-dimensional. 
For more on such examples see \cite{Lau}.

\vspace{3mm}
We will be interested in integrating forms on manifolds with 
corners. 
Integration over the boundary amounts to integrating 
over the boundary components.
We illustrate this with an example. 

\paragraph{Example 3. The square in $\R^2$.} 
The square is a manifold with corners of codimension-two.
Its edges are boundary hypersurfaces and its corners are 
codimension-two faces. 
Let $I\times I=[0,1] \times [0,1]$ be the unit square in $\R^2$,
 and 
suppose $\omega$ is a smooth 1-form on the boundary 
$\partial (I \times I)$. Consider the maps $F_i: I \to I\times I$ given by 
\( 
F_1(t) = (t,0)\;,
\quad
F_2(t) = (1,t)\;,
\quad
F_3(t) = (1-t,1)\;,
\quad
F_4(t) = (0, 1-t)\;\;.
\)
The four curve segments in the sequence traverse the 
boundary of $I \times I$ in the counterclockwise direction. 
Then Stokes' theorem for a manifold with corners gives 
\cite{Lee}
$
\int_{\partial (I \times I)} \omega=
\int_{F_1}\omega +
\int_{F_2}\omega +
\int_{F_3}\omega +
\int_{F_4}\omega$.
Such integration over rectangles should be familiar from electromagnetism, 
although it is usually not cast in this language. 
One of the main advantages of using manifolds with corners is that, 
for example, the cube which is not a smooth manifold would be smooth 
as a manifold with corners. 

\vspace{3mm} 
We will also need to study differential forms and cohomology on manifolds with corners.

\paragraph{$L^2$-cohomology.}
A manifold with corners can be viewed as a manifold with singularities.
De Rham cohomology does not capture the information at the singularities or corners.
To make up for this, one restricts to the subcomplex of square-integrable 
differential forms, which leads to $L^2$-cohomology. 
Let $(Y, g_Y)$ be a Riemannian manifold and let 
$\Omega^p=\Omega^p(Y)$ be the space of smooth $p$-forms and 
$L^2=L^2(Y)$ the $L^2$ completion of $\Omega^p$ with respect to the 
$L^2$-metric. The differential $d$ is defined to be the exterior differential 
with the domain 
$
{\rm dom}(d)= \{ \omega \in \Omega_{(2)}^p
: d\omega \in L^2(Y)\}
$,
where $\Omega_{(2)}^p=\Omega^p(Y) \cap L^2(Y)$
is the space of square-integrable smooth $p$-forms. 
The $L^2$-cohomology is 
then the cohomology of the cochain complex 
\(
0 \longrightarrow 
~\Omega_{(2)}^0(Y)~ \buildrel{d}\over{\longrightarrow} 
 ~ \Omega_{(2)}^1(Y)~ \buildrel{d}\over{\longrightarrow} 
 ~\Omega_{(2)}^2(Y)~ \buildrel{d}\over{\longrightarrow} 
 ~\Omega_{(2)}^3(Y)~ \buildrel{d}\over{\longrightarrow} 
\cdots\;,
\)
that is, 
$
H^p_{(2)}(Y)= \ker d_i/ {\rm Im}~d_{i-1}$.
The natural map
$
H_{(2)}^p (Y) \to H^p(Y;\R)
$
via the usual de Rham cohomology is an isomorphism for $Y$ 
a compact manifold with corners because the $L^2$ condition is 
automatically satisfied for all smooth forms. 
For a nice exposition on this see \cite{Da}.
Hodge theory for a manifold with corners is discussed in 
\cite{RS} \cite{Mu1}.

\paragraph{Smoothing corners.}
Manifolds with corners are smooth in the sense of having charts
locally as open subsets of $\R_+^n$. However, they look like they
should be singular at the corner. What is the explanation to this? 
One thing one could do is smooth out the corner via a diffeomorphism,
which is not an isometry. For instance, if the corner is that of a quadrant 
then one can replace rectangular coordinates with polar coordinates
and provide a smoothing of the corner by considering only nonzero 
value of the radial coordinate. This is called total boundary blow-up
\cite{MM} (see also \cite{Lo}). In our context will be interested in manifolds of the 
form $Z \cong [0,1)_{s_1} \times [0,1)_{s_2} \times M$, where 
$s_1$ and $s_2$ are Cartesian coordinates on the two intervals. 
Near the corner $M$,
introduce polar coordinates via $s_1=r \cos \theta$ and $s_2 = r \sin \theta$ so that
the totally blown-up space is $Z_{\rm tb} \cong [0, \varepsilon)_r 
\times [0, \pi/2]_\theta \times M$, for $\varepsilon >0$.
We have diffeomorphism instead of isometry because intersections of 
hypersurfaces at $M$ do not have to occur at right angles, but 
any angle in the plane can be related by a diffeomorphism to the 
standard upper right quadrant.
 We will have this blow-up
implicitly in mind in dealing with manifolds with corners.

\vspace{3mm}
To study the phase of the partition function we need to consider Dirac operators and 
their corresponding eta-invariants.

\paragraph{Continuous spectrum and the $b$-trace.}
 A Dirac operator on a manifold with corners 
has a continuous spectrum, and hence trying to define the 
 eta invariant will involve infinite traces. The way around this is to use 
 the $b$-trace within the $b$-calculus \cite{Mel}. For our purposes, 
 the main idea can be summarized as follows (see also the next example). 
 The $b$-trace is defined in terms of the $b$-integral for an operator $O$ (schematically) as 
\(
{}^{\bf b} {\rm Tr}(O):= ~{}^{\bf b}\hspace{-2mm} \int_Y {\rm tr} (O)\;.
\)
Then the corresponding eta-invariant will be defined using this trace as \cite{Mel}
\(
{}^{\bf b} \eta (D) =\frac{1}{\sqrt{\pi}} \int_0^\infty \frac{1}{\sqrt{t}}~ {}^{\bf b}{\rm Tr} (D e^{-tD^2}) dt\;.
\)
Let us illustrate this with an example, which will be useful for us later.

\paragraph{Example 4. The interval over a manifold.} 
Consider $[0,1]_s$, the 
unit interval with coordinate $s$, which will be fibered over 
a manifold $M$. 
The function $ds/s$ is not integrable over $[0,1]_s$, so that the 
corresponding heat operator is not trace class. However, the function 
$s^z$, ${\rm Re}z>0$, is integrable with respect to $ds/s$ over $[0,1]_s$. 
This suggests using an integral which corresponds to the usual integral when 
$z$ is zero. As nicely illustrated in \cite{Lo}, 
 let $f \in C^\infty (Y)$ be a smooth function on a manifold $Y$. 
 Then for all complex numbers $z$ with ${\rm Re}z>0$, the integral 
$F(z)=\int_Y s^z f dg$ exists and it extends from ${\rm Re}z>0$ to define a meromorphic 
function on all of $\C$. 
Note that $s^z=e^{z\log s}$ is an entire function of
$z$ for $s>0$. Thus, $f$ can be assumed to be supported on the collar 
$[0,1]_s \times M$ of $Y$. Then $F(z)$ is well-defined for ${\rm Re}z>0$
since $s^z f(s,m)$ is integrable 
with respect to the measure $(ds/s)dh$ as long as ${\rm Re}z>0$. 
Here  $m$ is a point in $M$ and $dh$ is a measure on $M$. 
Now expand 
$f(s,m)$ in Taylor series at $s=0$: $f(s,m) \sim \sum_{k=0}^\infty s^k f_k(m)$. Since
the integral
$\int_{[0,1]\times m} s^{z+k} f_k(m) \frac{ds}{s}dh$ is equal to  
$\frac{1}{z+k}\int_M f_k(m) dh$ then
the function $F(z)$ extends from ${\rm Re}z>0$ to be a meromorphic function on
$\C$ with only simple poles at $z=\{ 0, -1, -2, \cdots\}$ with residue at $z=0$ given by
$
\int_M f_0(m) dh= \int_M f(0,m) dh$.
The $b$-integral of $f$ is the regular value of $F(z)$ at $z=0$,
${}^{\bf b} \hspace{-1mm}\int_Y fdg= {\rm Reg}_{z=0} F(z)$, 
such that the residue of $F(z)$ at $z=0$ is given by 
$
{\rm Res}_{z=0} F(z)= \int_Y f(0,m) dh$.

\vspace{3mm}
We will also be interested in considering the kernels of Dirac operators on 
manifolds with corners. 

\paragraph{Infinite-dimensional kernels.}
The dimensions of the kernels
are generically infinite so that the Dirac operator is not Fredholm in general.
We will consider the effect of this in section \ref{sec dir}.
We now illustrate this in the simple example of the square. 
Consider the Cauchy-Riemann operator $\partial_z= \partial_x + i\partial_y$ on the square 
$[0,1]_x \times [0,1]_y$. The manifold and hence the operator are of product type. Then 
the kernel ${\rm ker} \partial_z$ is infinite-dimenisonal since this kernel  
consists of all holomorphic functions on the square.

\paragraph{Compactification of manifolds with cylindrical ends to manifold with corners.}
In order to deal with the non-Fredholm property of the Dirac operator on 
the manifold with corners $Z$, one has to introduce another manifold 
$\widehat{Z}$
of the same dimension
which is formed by attaching infinite cylinders to the collars of $Z$.
This will be used in section \ref{sec dir}.
The manifold $\widehat{Z}$ can be compactified by introducing 
the change of variables $x_1= e^{s_1}$ and $x_2=e^{s_2}$. As $s_i \to \infty$, $x_i \to 0$
and so this change of variables compactifies $\widehat{Z}$ to be the 
interior of a compact manifold with corners of codimension-two $Z$. 
The metric then transforms to the $b$-metric as
\(
g^Z= ds_1^2 + ds_2^2 + g^M~~ \leadsto~~ {}^{\bf b}g^Z= \left(\frac{dx_1}{x_1} \right)^2 + 
\left(\frac{dx_1}{x_1} \right)^2 + g^M\;.
\label{g to bg}
\)

\paragraph{The Maslov index.} 
When the twelve-manifold $Z^{12}$ has no boundary, there are nice additivity
properties, for example the Novikov additivity of the signature \cite{No}.
The Atiyah-Patodi-Singer index theorem \cite{APS}
gives the index of the Dirac and signature operators on manifolds with
boundary in terms of the $\widehat{A}$-genus and the L-genus, respectively,
and the  defects given by the corresponding eta invariants on the boundary.
In the case of corners, the signature is no longer additive, but there is a correction term
in Wall's nonadditivity \cite{Wa}. The signature defect is the Maslov
index of certain Lagrangian subspaces related to the cohomology of 
the boundary $Y^{11}$ and the corner. This, and the corresponding generalization 
using 
\cite{Bu} \cite{HMM} \cite{Mu} and \cite{Lo}, will be discussed in section 
\ref{sec dir} for the Dirac operator and in section \ref{sec sig}
for the signature operator. 

\subsection{Occurrence in M-theory}
\label{sec m}

In this section we consider four situations, three of which are related to M-branes in M-theory,
and one related to heterotic string theory, 
where manifolds with corners appear naturally. The first one is M-theory on AdS spaces,
which are 
configurations that occur as near horizon limits of M-branes. The second one arises by 
considering the M5-brane as a tubular neighborhood in eleven-dimensional spacetime
with boundary.
The third arises when considering boundaries in relation to the M2-brane. This 
includes the M2-brane having a boundary (ending on the M5-brane) or 
the M2-brane itself being considered as a boundary for instance when studying its partition function. 
The fourth views (a variant of) heterotic string theory as a corner in the twelve-dimensional 
bounding theory. 

\subsubsection{M-theory on AdS spaces}
\label{sec ads}

Anti-de Sitter space is a Lorentzian space with boundary at spatial infinity. 
The Euclidean version is given by a hyperbolic space, with very interesting
boundary structure at infinity. Therefore considering the boundary of AdS space
amounts, in an appropriate sense, to looking at the boundary of M-theory 
as $\partial {\rm AdS}_i \times S^{11-i}$ for $i=4,7$. 
Compactifying the Euclidean boundary gives a product of spheres.
In particular, for the M5-brane this gives $S^3 \times S^7$. 
Now the internal spaces $S^7$ and $S^4$ are boundaries of 
the 8-disk $\mathbb{D}^8$ and the 5-disk $\mathbb{D}^5$, respectively.
M-theory itself can be viewed as a boundary in twelve dimensions,
so that from the point of view of this bounding theory we have
spaces of the form ${\rm AdS}_i \times \mathbb{D}^{12-i}$ for $i=4,7$. 
We notice that both factors in the product are manifolds with boundaries, and hence the 
product itself is a manifold with corners of codimension-two, i.e. is a 
$\langle 2 \rangle$-manifold.

\vspace{3mm}
The internal spheres in the products with AdS spaces 
can also be replaced by homogeneous spaces $G/H$, 
where $G$ and $H$ are Lie groups,
with analogous near horizon structures \cite{CCDFFT}. 
In fact, general Einstein spaces $M^{11-i}$, for $i=4,7$,
with Killing spinors -- and hence are Spin --
can be used as well (see \cite{Fi} and references therein). 
Thus, in order to detect corners, we would like to ask whether the spaces 
${\rm AdS}_i \times M^{11-i}$ can be lifted to twelve 
dimensions. This reduces to checking whether 
$M^{11-i}$ can be boundaries. For $M^7$ this is 
always the case since the relevant bordism group
is trivial $\Omega_7^{\rm Spin}=0$; that is the Spin manifold 
$M^7$ is always the boundary of some eight-manifold, 
say $W^8$. However, for $M^4$ this is not the case since
the bordism group is not trivial, $\Omega_4^{\rm Spin}=\Z$.
By Rohlin's theorem, a closed oriented Spin 4-manifold 
$M^4$ is null cobordant in $\Omega_4^{\rm Spin}$, i.e. is the
boundary of a compact oriented Spin smooth 5-manifold 
$W^5$ if and only if the signature $\sigma (M^4)$ of $M^4$ vanishes. 
Thus the signature is a complete cobordism invariant. 
The 
isomorphism $\Omega_4^{\rm Spin} \buildrel{\cong}\over{\to} \Z$
sends any cobordism class $[M^4]$ to $\sigma(M^4)/16$.
 In particular, the Kummer 
 surface $K_4=\{ z_1^4 + z_2^4 + z_3^4 + z_4^4\} 
 \subset \C P^3$, whose signature is $\sigma (K_4)= -16$, provides a generator 
 for $\Omega_4^{\rm Spin}$. We have

\begin{proposition}
$(i)$ The near horizon limit of the M2-brane can always be described as corner
for the twelve-dimenisonal bounding theory. 

\noindent $(ii)$ The near horizon limit of the M5-brane can can be described as a corner 
for the twelve-dimensional bounding theory provided that the internal 
four-manifold is an Einstein space with zero signature.  
\label{prop nhl}
\end{proposition}

We will consider the M5-brane and the M2-brane themselves 
in section \ref{sec M5} and section \ref{sec M2}, respectively.

\vspace{3mm}
 Examples of Spin 4-manifolds with zero signature include 
 the 4-sphere $S^4$, the projective space 
 $\R P^4$ and their quotients by finite groups. 
 Dimensional reductions of the latter
 type are considered e.g. in \cite{FS}.
 Classes of examples include
 ones for which the $\widehat{A}$-genus vanishes, since in four
 dimensions the $\widehat{A}$-genus and the Hirzebruch $L$-genus
  are related by a simple numerical factor. By the result of Atiyah-Hirzebruch 
the $\widehat{A}$-genus vanishes if the manifold admits 
 a smooth (isometric) circle action \cite{AH}. Interestingly, even in the 
 non-Spin case (say for us Spin${}^c$), such a result still holds
 \cite{HH}. The resulting theory on the orbit of the circle action 
 is ten-dimensional type IIA string theory. 
 The M5-brane will give rise to a type IIA NS5-brane, which is of the same
 dimension, so that the dimension of the transverse space is reduced by one.
 Therefore, we have the
 nice compatibility result

\begin{proposition}
The near horizon limit of the M5-brane can be described as a corner
when M-theory is taken 
with a circle action, that is when the theory is related to type IIA string theory.
\end{proposition}

For example, for $M^4=S^3 \times S^1$ this leads to type IIA string theory on 
${\rm AdS}_7 \times S^3$, studied e.g. in \cite{SS}.
On the other hand, for $S^7$ the circle action
 gives a supersymmetric background in type IIA string theory
of the form ${\rm AdS}_4 \times \C P^3$ first considered in \cite{DLP}.
In these cases the ten-dimensional 
corners are $\partial {\rm AdS}_7 \times S^3 \times S^1$
and $\partial {\rm AdS}_4 \times S^7$, respectively.

\subsubsection{The M5-brane as a tubular neighborhood}
\label{sec M5}

 Here we consider the extension of the description of the
 M5-brane as a tubular neighborhood  
  to the case when $Y^{11}$ has a boundary.
 This results, upon removing of a tubular neighborhood, in 
 a manifold with corners of codimension-two.
 
\vspace{3mm}
Consider an M5-brane with worldvolume $W^6$, considered as a 
(closed) submanifold inside 
 a closed eleven-manifold $Y^{11}$. Removing a tubular neighborhood
 of the M5-brane leads to a manifold with a boundary, as illustrated in 
 \cite{DFM} and used in \cite{DMW-boundary}. While both of these references
 are concerned mainly with the case when $Y^{11}$ has a boundary, 
 that was restricted to a closed $Y^{11}$
 when dealing with tubular neighborhoods. 
 Now we provide a description of the case when $\partial Y^{11}\neq \emptyset$
 using the formalism in \cite{Ce} \cite{Do} \cite{Ja}.

\vspace{3mm}
Let $\iota : W^6 \hookrightarrow Y^{11}$ be the embedding of the M5-brane in 
spacetime with normal bundle $N^{11} \to W^6$, viewed as a tubular neighborhood of 
$W^6$ in $Y^{11}$. The unit sphere bundle of radius $r$ is the 
associated bundle $S^4 \to \cS^{10} \to W^6$, and the corresponding 
disk bundle of radius $r$ is $\mathbb{D}^5 \to \D^{11} \to W^6$.
Removing this disk bundle leads to an eleven-manifold
$Y^{11}_r=Y^{11} - \D^{11}$ with boundary $\partial Y^{11}_r=\cS^{10}$,
the sphere bundle.

\vspace{3mm}
If $Y^{11}$ is a manifold with boundary then the removal of a tubular 
neighborhood of the M5-brane from $Y^{11}$ will result in 
a manifold with corners of codimension-two. 
Then,  assuming that $Y^{11}$ has multiple boundary components
$\partial_i Y^{11}$ ($i=1, \cdots, n$), 
$W^6$ is a manifold with faces and becomes a manifold with 
boundary if we identify $\partial_i W^6= W^6 \cap \partial_i Y^{11}$. 
We can interpret this as the boundary of M5-brane on the M9-brane,
or the M5-brane in heterotic M-theory. 

\vspace{3mm}
Let us now consider the relation to type IIA string theory. For that, we 
assume that $Y^{11}$ admits a differentiable circle action as in 
\cite{DMW} \cite{MS} \cite{S-gerbe}, and 
assume that the boundary $\partial_i Y^{11}$
is invariant under this circle action. 
We would like to identify the corner in this case. 
The set of nonzero normal vectors $\{ v\}$ is $N^{11}-W^6$. This is acted upon
by the positive real line 
$\R^*_+=\{ \alpha \in \R ~|~\alpha >0\}$ via multiplication by a positive scalar:
$v \mapsto \alpha v$.
The sphere bundle $\cS^{10}$
can be identified with the quotient  $(N-W^6)/\R^*_+$.
We extend to the cylinder bundle over the sphere bundle 
$\cS^{10} \times \R=\cC^{11}$,
which is a trivial line bundle
the fiber at $\R^*_+ v$ being 
$\R v$, and identify $\cS^{10}$ with the zero section of $\cC^{11}$. 
Let $\cC^{11}_+ \subset \cC^{11}$ 
be the non-negative half of $\cC^{11}$; an element 
$rv$ of the fiber of $\cC^{11}$ over $\R^*_+ v$ is 
in $\cC_+^{11}$ if $r \geq 0$. 
If $U$ is an open neighborhood of $W^6$ in $N^{11}$, denote by 
$\cC_+U$ the inverse image of $U$ under the canonical map
$\cC_+^{11} \to N^{11}$.

\vspace{3mm}
The $S^1$-manifold with corners of codimension-two
will be, as a set, the disjoint union $(Y^{11}-W^6)\cup \cS^{10}$.
Define a tubular neighborhood map, that is an $S^1$-equivariant 
diffeomorphism $T$ of an open $S^1$-invariant 
neighborhood $U$ of $W^6$ in $N^{11}$ onto an open neighborhood
$U'$ of $W^6$ in $Y^{11}$ with the properties that 
$T|_{W^6}$ is the inclusion map $W^6 \subset Y^{11}$ and
 the induced map $T_*: N^{11} \to N^{11}$ of the normal bundle of
$W^6$ in $U$ into the normal bundle of $W^6$ in $U'$ is the 
identity map. 
The tubular map (see \cite{Do}) induces a map $T': \cC_+U \to (Y^{11}-W^6) \cup \cS^{10}$,
with respect to which $T'$ is a diffeomorphism onto a neighborhood 
of $\cS^{10}$ and which induces the given structure on $(Y^{11}-W^6)$.
Now let $T_1$ be a second tubular map, thus defining 
a second structure on $(Y^{11}-W^6)\cup \cS^{10}$. Then the identity map
on $(Y^{11}- W^6) \cup \cS^{10}$ is an isomorphism of the these two
structures, so that $(Y^{11}-W^6)\cup \cS^{10}$ becomes 
a well-defined manifold with corners. 

\vspace{3mm}
Let $p: (Y^{11}- W^6)\cup \cS^{10} \to Y^{11}$ be the natural 
projection, which is the identity on $(Y^{11}- W^6)$ and 
bundle projection on $\cS^{10}$. Define the boundary to be 
$\partial_i( (Y^{11}- W^6)\cup \cS^{10})=p^{-1} (\partial_i Y^{11})$
for $i=0,1$ 
\(
\xymatrix{
(Y^{11}- W^6)\cup \cS^{10}
\ar[d]^{{\rm id}\times {\rm pr}}
~
&
~~
\partial_i\left(  (Y^{11}- W^6)\cup \cS^{10} \right)
\ar@{_{(}->}[l]
\ar[d]^p\\
Y^{11}
~
&
~
\partial_iY^{11}
\ar@{_{(}->}[l]
}
\)
and $\partial_2 ( (Y^{11}- W^6)\cup \cS^{10})=\cS^{10}$. 
This shows that $((Y^{11}- W^6)\cup \cS^{10})$ becomes an
$S^1$-manifold with corners of codimension-two. 
We summarize

\begin{proposition}
The M5-brane worldvolume in an eleven-dimensional manifold with 
boundary is a manifold with corners, 
described above.
\end{proposition}

\subsubsection{The M2-brane and boundaries}
\label{sec M2}

Consider M-theory on a Spin eleven-manifold $Y^{11}$ which is a product
of two Spin manifolds
$X^3 \times M^8$. Take $X^3$ to be a three-manifold with a boundary
$\partial X^3=\Sigma_g$, a Riemann surface, and $M^8$ a closed 
eight-manifold. Now take  
an M2-brane with boundary to wrap around $X^3$ and identify the 
boundary of the M2-brane with the boundary of $X^3$. Then we try 
to lift to twelve dimensions by making $M^8$ into a boundary of
a nine-dimensional manifold $N^9$, with $\partial N^9=M^8$. 
However, we cannot always perform these steps
because the
Spin cobordism group in eight dimensions is not zero. In fact,
$\Omega_8^{\rm Spin}\cong \Z \oplus \Z$, generated by the 
quaternionic projective plane $\mathbb{H}P^2$ and 
a generator which is one-fourth the square of the Kummer surface
$\frac{1}{4}(K3)^2$.
If we were to always find a Spin boundary
 then we would consider $\Sigma_g \times M^8$ is the corner
of the twelve-dimensional manifold $Z^{12}$. The latter is the product of 
two manifolds with boundary, namely $X^3$ and $N^9$, and so indeed 
it is a manifold with corners of codimension-two.

\vspace{3mm}
We could also try to take $Z^{12}$ to be just oriented and not necessarily 
Spin, and the same for $M^8$. Then in trying to lift from $M^8$ to $N^9$
(again just oriented) we have to check that the obstruction in the 
oriented cobordism group in dimension eight is zero. In general this is 
not the case since $\Omega_8 \cong \Z \oplus \Z$, generated by the 
projective spaces $\CP^4$ and $\CP^2 \times \CP^2$.
We then have

\begin{proposition}
The M2-brane with a boundary gives rise to a ten-dimensional corner
in the twelve-dimensional theory, provided an eight-dimensional 
zero bordism is used for the transverse space.
\end{proposition}

\subsubsection{The heterotic theory as a corner}

The topological and global analytic aspects of M-theory are best 
described using a lift from eleven dimensions to twelve dimensions,
where the theory on $Y^{11}$ is considered from the point of view of the
theory on a twelve-dimensional Spin manifold $Z^{12}$ bounding 
$Y^{11}$, that is $\partial Z^{12}=Y^{11}$ \cite{Flux}. On the other hand,
heterotic string theory can be considered on M-theory with a boundary, that
is when $Y^{11}$ itself has a boundary. The naive boundary of a boundary 
does not exist. However, manifolds with corners come to the rescue, so 
that heterotic string theory can be viewed as a corner of the 
twelve-dimensional theory and the picture is consistent.

\vspace{3mm}
The rest of the paper will concerned with expanding around this 
interpretation.
In the following section we will consider analytical and geometric
consequences of viewing the heterotic 
theory as a corner. 

\section{Analytical and geometric aspects of M-theory with corners}
\label{sec anal}

The goal of this section is to explore analytical consequences of taking 
the heterotic theory to be a corner in the twelve-dimensional theory. 
Our discussion will mostly focus on the Dirac and signature 
operators, their
eta-invariants, and the corresponding phase of the partition function.

\subsection{Formulation using Dirac operators}
\label{sec dir}
In this section we consider M-theory on two disconnected components, both of 
which form the boundary of the twelve-dimensional bounding theory, and 
which intersect on one corner, representing the heterotic theory.  
This is the opposite to the usual situation, where M-theory is taken on one
component and the heterotic theory is taken on two disconnected components.
The analytical constructions we apply here are very nicely surveyed in 
\cite{Lo}, to which we refer heavily throughout this section.
Mass regularizations and perturbations will play an important role.

\paragraph{The fields in heterotic string theory.} 
 Let $S^\pm$ denote the Spin bundles on $M^{10}$. 
The fermionic fields in heterotic string theory consist of a gravitino
$\psi$, which is a section of $T^*M^{10} \otimes S^+$, 
 a dilatino $\lambda$, which is a $\frak{e}_8$-valued 
 section of $S^+$, and 
a
gaugino $\chi$, which is a section of $S^-$.
Here $\frak{e}_8$ is the Lie algebra of the Lie group $E_8$. 
We will work with general twisted spinors, that is with sections of
$S^\pm \otimes E$, where the vector bundle $E$ 
can be taken as the $E_8$ bundle or the tangent bundle 
(minus appropriate number of trivial line bundles) according to 
the context.

\paragraph{The case with no corners.}
For comparison, let us briefly recall the case with no corners \cite{F}.
Consider $Y^{11}=[0,1] \times M^{10}$ with the product metric, 
where $M^{10}$ is a closed ten-dimensional Spin manifold. 
Then $\partial Y^{11}= M_0 \cup M_1$, where 
$M_1 \cong M^{10}$ and $M_0 \cong - M^{10}$ (that is,
$M^{10}$ with the opposite
orientation).  Let $P^\pm$ be the local boundary conditions for the 
Dirac operator $D_Y$ corresponding to spinors in $S^\pm_{\partial Y}$ 
being zero, imposed respectively on $M_0$ and $M_1$. 
Then \cite{F}
\(
{\rm index} (D_Y, P^\pm)= {\rm index}(D_M)\;,
\label{fr}
\)
where $D_M$ is the Dirac operator on $M^{10}$. 
As explained in \cite{F}, this is the case for Horava-Witten theory
\cite{HW1} \cite{HW2}, which we consider in more detail at the end of this section.
The heterotic theory can also 
be viewed from the point of view of reduction of M-theory on 
$S^1/\Z_2$, where $\Z_2$ is an orientation-reversing involution.
More generally, let $Y^{11}$ be an eleven-manifold with an orientation-reversing 
isometric involution $\tau: Y^{11} \to Y^{11}$ and with a lift $\widetilde{\tau}: SY^{11} 
\to SY^{11}$ to the Spin bundle which anticommutes with the 
Dirac operator $D_Y$ and satisfies $\widetilde{\tau}^2=1$. Then 
$
D_Y: S^\pm Y^{11} \longmapsto S^\mp Y^{11}$,
where $S^\pm Y^{11}$ are the $\pm$-eigenspaces of $\widetilde{\tau}$. 
When $Y^{11}= S^1 \times M^{10}$ with $\tau$ a reflection on the 
circle $S^1$ and $M^{10}$ is a compact Spin ten-manifold, 
then the same formula \eqref{fr} holds \cite{F}.

\paragraph{The case with corners.}
Now consider $Z^{12}$ as a compact oriented Riemannian 12-manifold with corners 
of codimension-two and metric $g^Z$. 
Assume that $Z^{12}$ has exactly two boundary hypersurfaces $Y^{11}_1$ and $Y^{11}_2$
that intersect in exactly one codimension-two face $M^{10}$. 
The two hypersurfaces correspond to M-theory on two disconnected spaces. 
Near each hypersurface $Y^{11}_i$, we assume that $Z^{12}$ has a collar
neighborhood $Z^{12}\cong [0,1)_{s_i} \times Y^{11}_i$ where the 
metric is a product $g^Z= ds_i^2 + g^Y_i$, with $g_i^Y$
 the metric on $Y_i^{11}$.
Then the product decomposition near each $Y_i^{11}$ can be taken to 
be $Z^{12}\cong [0,1)_{s_1} \times [0,1)_{s_2} \times M^{10}$ near the 
corner where the metric is a product 
$g^Z= ds_1^2 + ds_2^2 + g^M$, with $g^M$ a metric on $M^{10}$. 
Here $s_1$ and $s_2$ are, as before,
 the coordinates on the `square' over $M^{10}$. 
 In what follows we apply some of the results (surveyed) in \cite{Lo}.

\paragraph{The resulting Dirac operators on $Y^{11}_i$ and on 
$M^{10}$ starting from one on $Z^{12}$.} 
Let $E$ and $F$ be Hermitian vector bundles over $Z^{12}$
whose restrictions to $Y^{11}_i$ are $E_i$ and $F_i$, $i=1,2$,
respectively. 
The restrictions to $M^{10}$ are denoted $E_0$ and $F_0$. 
What we have in mind are Spin bundles, possibly twisted by vector bundles, 
such as an $E_8$ vector bundle or the tangent bundle. Starting with 
a Spin bundle $S_Z=S_Z^+ \oplus S_Z^-$ on $Z^{12}$, this reduces
to $S_Y=S^+_Z$ or $S_Y=S^-_Z$ on $Y^{11}$. The choice depends on the 
boundary conditions. 
In turn, the restriction of $S_Y$ to the ten-dimensional boundary will be 
$S_Y|_{M^{10}}\cong S_M \cong S^+_M \oplus S_M^-$. The two splittings 
lead to local boundary conditions for the Dirac operators  on $Z^{12}$ and on 
$Y^{11}$.

\vspace{3mm}
Let 
$
D: C^\infty (Z^{12}, E) \to C^\infty (Z^{12}, F) 
$
be a Dirac operator on $Z^{12}$ which 
is of product type 
$
D= \Gamma_i( \partial_{s_i} + D_i)
$
near each hypersurface
on the collar $Z^{12} \cong [0,1){s_i} \times Y^{11}_i$, where $\Gamma_i$ is a 
Dirac matrix, i.e. 
a unitary isomorphism from $E_i$ into $F_i$ and where 
$
D_i: C^\infty (Y_i^{11}, E_i) \to C^\infty (Y_i^{11}, E_i) 
$
is a (formally) self-adjoint Dirac operator on the 11-dimensional manifold with 
boundary $Y^{11}_i$. 
Furthermore, assume that on the product decomposition near the corner, the 
Dirac operator takes the form
$
D= \Gamma_1 \partial_{s_1} + \Gamma_2 \partial_{s_2} + B
$
where 
$
B : C^\infty (M^{10}, E_0) \to C^\infty (M^{10}, F_0)
$
is a Dirac operator on the ten-dimensional manifold without 
boundary $M^{10}$. 
On the collar $Z^{12}\cong [0,1)_{s_1} \times [0,1)_{s_2}\times M^{10}$
we have 
$
\Gamma_i (\partial_{s_i} + D_i) = \Gamma_i \partial_{s_i} + B$, $(i=1,2)$,
so that 
\(
D_1= \Gamma_1^{-1} \Gamma_2 \partial_{s_2} + \Gamma_1^{-1} B
~~~{\rm and}~~~
D_2= \Gamma_2^{-1} \Gamma_1 \partial_{s_1} + \Gamma_2^{-1} B
\label{eq 2 Dirac}
\)
The fact that each $D_i$ is (formally) self-adjoint, $D_i^*=D_i$ is compatible
with the Clifford algebra identity 
$
\Gamma_i^{-1} \Gamma_j + \Gamma_j^{-1} \Gamma_i = 2\delta_{ij}$ and gives
the condition
$B^* \Gamma_i= \Gamma_i^{-1} B$(
here $\Gamma^{-1}=\Gamma^*$).
Then we relate the 11-dimensional operator to the 10-dimensional operator via
$
D_1= \Gamma (\partial_{s_2} + D_M)$, where
$\Gamma= \Gamma_1^{-1} \Gamma_2$ and $D_M= \Gamma_2^{-1}B$. 
the operator $D_M$ is the Dirac operator on $M^{10}$ induced by $D_1$. 
The Dirac operator induced from $D_2$ has a simple expression
in relation to $D_M$ and hence can be considered equivalent:
$
D_2= - \Gamma (\partial_{s_1} + \widetilde{D}_M)$, with
$\widetilde{D}_M= \Gamma D_M$.

\vspace{3mm}
Since $\Gamma^2=-{\rm Id}$ then $\Gamma: E_0 \to E_0$ 
has eigenvalues $\pm i$. Let $E_0^\pm$ denote the eigenspaces 
corresponding to the eigenvalues $\pm i$. These are subbundles of 
$E_0$ and 
\(
E_0= E_0^+ \oplus E_0^-
\label{z2 gr}
\)
is an orthogonal decomposition since $\Gamma$ is unitary. 
Furthermore, $D_M$ is odd with respect to $\Gamma$: $D_M \Gamma=-\Gamma D_M$, so 
$D_M$ is also odd with respect to the $\Z_2$-grading 
\eqref{z2 gr}. Therefore,

\begin{lemma}
The Dirac operator on the ten-dimenisonal heterotic corner $M^{10}$ induced from 
twelve dimensions via $D= \Gamma_1 \partial_{s_1} + \Gamma_2 \partial_{s_2}
+ \Gamma_2 D_M$ 
takes the form 
\(
D_M= 
\left[
\begin{array}{cc}
0 & D_M^-\\
D_M^+ & 0
\end{array}
\right]
~:~
C^\infty (M^{10}, E_0^+ \oplus E_0^-)
\to
C^\infty (M^{10}, E_0^+ \oplus E_0^-)\;,
\)
where $D_M^\pm$ are the restrictions of $D_M$ to $C^\infty (M^{10}, E_0^\pm)$.
Self-adjointness of $D_M$ implies that $(D_M^+)^*=D_M^-$. 
\end{lemma}

Here $E_0^\pm$ is $S_M^\pm$ twisted with the $E_8$ vector bundle.
 Note that $D_M^+$ is the operator appearing in Horava-Witten theory \cite{HW1}
 \cite{HW2}.

\paragraph{Square-integrability and Sobolev spaces.}
We have seen in section \ref{sec cor} that manifolds with corners require working with
square integrable differential forms. 
Denote the Sobolev space of order $k$ by $\cH^k$. 
Thus, for $E$ a vector bundle as above, $\cH^k(M^{10}, E)$ denotes the $E$-spinors on $M^{10}$ 
for which $(D_M)^j \psi$, $j=0, \cdots, k$, is square-integrable. 
So $\cH^1(M^{10}, E)$ is the natural domain for $D_M$. 
From the Atiyah-Singer index theorem, the index of the 
Dirac operator 
\(
D_M^+: \cH^1(M^{10}, E_0^+) \to L^2(M^{10}, E_0^-)
\label{eq DM}
\)
is zero, ${\rm Ind} D_M^+=0$, as in \cite{F}. Since 
$(D_M^+)^*=D_M^-$, it follows that dim ker$D_M^+=$dim ker$D_M^-$.
Now consider the Dirac operator 
 $D: \cH^1(Z^{12}, E) \to L^2(Z^{12}, F)$
on $Z^{12}$. This is never 
Fredholm as dim ker$D=\infty$. Therefore, we need to replace this operator 
by another (in a sense equivalent) Dirac operator which is Fredholm.

\vspace{3mm}
Let $\widehat{Z}^{12}$ be the manifold formed by taking the infinite 
cylinder $(-\infty, 0]_{s_1} \times Y_1^{11}$
and attaching it to the collar $[0,1)_{s_1} \times Y^{11}_1$
of $Z^{12}$, then taking  $(-\infty, 0]_{s_2} \times Y_2^{11}$
and attaching it to the collar $[0,1)_{s_2} \times Y^{11}_2$, and finally
taking $(-\infty, 0]_{s_1} \times (-\infty, 0]_{s_2} \times M^{10}$ and 
attaching it to the remaining open quadrant. 
Since all geometric structures and the Dirac operator
are of product type near the boundary of $Z^{12}$, they all have natural 
extensions to the manifold $\widehat{Z}^{12}$. Let $\widehat{D}$
be the extension of the Dirac operator $D$ to $\widehat{Z}^{12}$.
When attaching ends to a manifold, one also talks about weighted Sobolev spaces
which arise by considering 
the weighted (or conformal) Dirac operator 
$e^{-\alpha s} \widehat{D} e^{\alpha s}$, where 
$s$ is a coordinate function on the cylindrical end and $\alpha$ is a constant 
whose absolute value is less than the smallest absolute value of 
a nonzero eigenvalue of $D_M$. From 
$
e^{-\alpha s} \widehat{D} e^{\alpha s}= D + \alpha \Gamma
$
we see that $\alpha$ plays the role of mass for the spinors.  
Then $e^{-\alpha s} \widehat{D} e^{\alpha s}: \cH^1(\widehat{Z}^{12}, E) \to L^2(\widehat{Z}^{12}, F)$
can be replaced, for $|\alpha|>0$ sufficiently small, by 
\(
\widehat{D}: e^{\alpha s} \cH^1(\widehat{Z}^{12}, E) \to e^{\alpha s} L^2(\widehat{Z}^{12}, F)\;.
\)
The variable $s$ can be replaced with the variable $x=e^s$. 
We will shortly use two variables, one for each interval (see expression \eqref{eq wss}). 
We summarize the conditions on the spinors

\begin{lemma}
$(i)$ Both the spinor $\psi$ and the mass-normalized spinor 
$m\psi$ have to be integrable
on $\widehat{Z}^{12}$.

\noindent $(ii)$  The normalized spinors on $\widehat{Z}^{12}$ vanish as $s \to \infty$
and coincide with the spinors when $s=0$. 
\label{lemma both}
\end{lemma}

Below we consider conditions, coming from the corner, for when $\widehat{D}$ is 
Fredholm. 
We will consider two cases, according to whether or not the corner Dirac operator is
invertible. 

\subsubsection{The non-supersymmetric case}

In this section we assume 
that the corner Dirac operator $D_M$ is invertible. 
Consider $\widehat{Y}_i^{11}$, the eleven-manifold with cylindrical end
formed by attaching an infinite cylinder to the 
eleven-dimensional compact manifold with boundary $Y^{11}_i$.
This has infinite volume so that the spectrum of the corresponding 
Dirac operator $\widehat{D}_i$ is continuous rather than discrete. 
Then the eta-invariant cannot be defined since it involves a trace. 
As explained in section 
\ref{sec base},
the way around this is to use a $b$-trace.

\vspace{3mm}
Unlike the case of a manifold with boundary, on a manifold with corners 
the Dirac operator on $Z^{12}$ cannot always be made into a Fredholm 
operator. In fact, \cite{LM} there exists a $\delta >0$ such that for all
$0 < |\alpha_i| ,\delta$, $i=1,2$, the Dirac operator on weighted Sobolev spaces
\(
\widehat{D}: e^{\alpha_1 s_1} e^{\alpha_2 s_2} 
\cH^1(\widehat{Z}^{12}, E) \to e^{\alpha s_1} e^{\alpha_2 s_2}
 L^2(\widehat{Z}^{12}, F)
\label{eq wss}
\)
is Fredholm if and only if the corner operator 
$D_M: \cH^1(M^{10}, E_0) \to L^2(M^{10}, E_0)$
is invertible (has zero kernel). Generally,
a Dirac operator on a noncompact manifold is Fredholm 
if and only if it is invertible `at infinity'.  
That is \cite{Mu} (also see \cite{LM})
the Dirac operator 
$
\widehat{D}: \cH^1(\widehat{Z}^{12}, E) \to L^2(\widehat{Z}^{12}, F)
$
is Fredholm if and only if $\widehat{D}_i: \cH^1(\widehat{Y}^{11}_i, E_i) 
\to L^2(\widehat{Y}_i^{11}, E_i)$ for $i=1,2$, and the corner operator 
$D_M: \cH^1(M^{10}, E_0) \to L^2(M^{10}, E_0)$ are
each invertible. 
This places conditions on the topology of 
$M^{10}$.

\vspace{3mm} 
Let us consider the index of the Dirac operator coupled to an $E_8$ bundle $E$ on
the heterotic corner $M^{10}$.  
The integrand in the index is 
\(
\widehat{A}(M^{10}) {\rm ch}(E)= c_1(E) \widehat{A}_2 + {\rm ch}_3(E) \widehat{A}_1 + {\rm ch}_5(E)\;.
\)
This is automatically zero for $E_8$ since in this case ${\rm ch}_i(E)=0$, $i=1,3,5$.
Note that the characteristic classes of $E$ are all in dimensions divisible by 4,
and are given by $248 + 60 a + 6a^2 + \frac{1}{3}a^3$, where $a$ is the degree four 
class characterizing the bundle. Since ${\rm Index}(D_M)=\dim\ker D_M
- \dim{\rm coker}D_M$, then the vanishing of the both the index and the 
dimension of the kernel gives that the cokernel is also trivial. Therefore, we have
 
\begin{proposition}
Requiring the Dirac operator in twelve dimensions to be Fredholm is equivalent 
to
the Dirac operator on the heterotic corner being invertible, i.e. having zero kernel.
This results in a non-supersymmetric theory. Furthermore, in this case  the 
cokernel is also zero. 
 \end{proposition}

An example of a non-supersymmetric heterotic theory is the model given in \cite{FH}.

\vspace{3mm}
We have seen (cf. just before Lemma \ref{lemma both})
that $\widehat{Z}^{12}$ can be transformed to $Z^{12}$,
with the metric transforming as in \eqref{g to bg}.
Similarly, the Dirac operator transforms to the $b$-Dirac operator as
\(
\widehat{D}=\Gamma_1 \partial_{s_1} + 
\Gamma_2 \partial_{s_2} + B ~~\leadsto~~ 
{}^{\bf b}\widehat{D}=\Gamma_1 x_1\partial_{x_1} + 
\Gamma_2 \partial_{x_2} + B\;,
\)
which acts as ${}^{\bf b}\widehat{D}: x^\alpha \cH^1(Z^{12}, E) \to x^\alpha L^2_b(Z^{12}, F)$,
where $\cH^1_b$ and $L^2_b$ denote $b$-Sobolev spaces and 
space of square integrable functions using the $b$-integral. 
This operator is Fredholm if and only if the corner operator 
$D_M$ is 
invertible (has zero kernel) \cite{Mu}.


\vspace{3mm}
Note that for general boundaries -- even without corners -- 
the eta-invariant $\eta (\widehat{D}_i)$ 
should in general be replaced with the $b$-eta-invariant 
${}^{\bf b}\eta (\widehat{D}_i)$ via replacing the trace Tr with the $b$-trace
${}^{\bf b}{\rm Tr}$. In this case, the APS index theorem 
in the setting above becomes \cite{Mu} 
\(
{\rm ind}_\alpha \widehat{D} =\int_{Z^{12}} 
\widehat{A}(Z^{12}) {\rm ch}(E)
 -\frac{1}{2} \sum_{i=1,2}
\left\{
{}^{\bf b}\eta(\widehat{D}_i) + {\rm sign}\alpha \cdot \dim \ker\widehat{D}_i
\right\}\;,
\label{eq mmu}
\)
where $\alpha=(\alpha_1, \alpha_2)$.  Then we can reformulate the phase of the partition
function using the $b$-eta-invariant by specifying the vector bundles $E$ and $F$ 
to the $E_8$ bundle and to the Rarita-Schwinger bundle. Hence,

\begin{proposition}
The phase of the partition function for general boundary conditions 
for the eleven-dimensional boundary is
$\exp 2\pi i \left[ \frac{1}{4} {}^{\bf b}{\overline{\eta}}_{E_8} 
+ \frac{1}{8}{}^{\bf b}{\overline{\eta}}_{RS} \right]$.
\label{prop etabar}
\end{proposition}

\paragraph{Remark on the number of zero modes.}
Note that expression \eqref{eq mmu} can be rewritten in terms of $b$-calculus
by using more general boundary conditions than the one in the original 
 Atiyah-Patodi-Singer treatment \cite{APS};
they are called augmented APS boundary conditions \cite{HMM}. 
This way the number of zero modes, i.e. the dimensions of kernels of the Dirac
operators, are absorbed into the index. The number of zero modes
taken mod 2, that is the mod 2 index of the $E_8$ Dirac operator in ten dimensions, 
plays a crucial and extensive role in the discussions in \cite{DMW}, which use
the APS boundary conditions. Then we can see that when we use augmented
boundary conditions, these zero mode terms are `absent', and hence
presumably cannot detect an anomaly. Therefore, we see that the use of 
more general boundary conditions via $b$-calculus seems to 
drastically simplify
the discussion in \cite{DMW}.

\paragraph{Extension to more corners.}
The above results still hold if $Z^{12}$ has more than one corner provided that 
each corner Dirac operator has zero index.
This allows the construction of a separate perturbation for each corner
\cite{LM}.
The assumption can be removed by including a larger class of perturbations 
called `overblown' $b$-smoothing operators
\cite{Lo}.

\subsubsection{The supersymmetric case}

We now consider the case when the corner Dirac operator not invertible, that is
the Dirac operator has a nonzero kernel and so there are
zero modes for the spinors, as appropriate for a supersymmetric theory. 
Dropping the invertibility assumption on the corner Dirac operator $D_M^+$
requires the use of perturbations (see \cite{Lo} for a description  of the formalism). 
For comparison with the boundary-only case, see \cite{F}.
The operator \eqref{eq DM}
has zero index, that is $\dim\ker D_M^+=\dim\ker D_M^-$. 
The kernel is exactly the obstruction to $\widehat{D}$ being a Fredholm
operator on weighted Sobolev spaces, so the perturbations are chosen 
to be isomorphisms on the kernel. Let $T: {\rm ker}D_M \to {\rm ker}D_M$
be a self-adjoint unitary isomorphism that anticommutes with
 $\Gamma=\Gamma_1^{-1}\Gamma_2$ (see expressions \eqref{eq 2 Dirac}), 
 so $T$ decomposes as an odd matrix
\(
T=
\left[
\begin{array}{cc}
0 & T^-\\
T^+ & 0
\end{array}
\right]~
: ~
{\rm ker}D_M^+ \oplus {\rm ker}D_M^- \to
{\rm ker}D_M^+ \oplus {\rm ker}D_M^-\;,
\)
where $T^\pm: {\rm ker} D_M^\pm \to {\rm ker}D_M^\mp$ are unitary 
isomorphisms with respect to the inner product on 
${\rm ker} D_M \subset L^2(M^{10}, E_0)$. 

\vspace{3mm}
Let $\{ \psi^+_j\}_{j=1}^N$
and $\{ \psi^{-}_j\}_{j=1}^N$ be spinor orthonormal bases of
${\rm ker}D_M^+$ and ${\rm ker}D_M^-$, respectively, with
$\psi_j^+, \psi_j^{-} \in C^\infty(M^{10}, E_0)$
spinors on $M^{10}$, possibly twisted with the tangent bundle or 
with the 
$E_8$ vector bundle. 
Then $T$ has the expression in terms of fermion bilinears
\(
T= \sum_{j=1}^N \psi^+_j  \otimes \overline{\psi}_j^{-} 
+
 \sum_{j=1}^N \psi_j^{-}  \otimes \overline{\psi}_j^+\;.
\)
and
is a smoothing operator on $M^{10}$, that is $T$ has a smooth kernel.
Since
$\ker D_M^+$ is a dual vector space to $\ker D_M^-$,
we interpret $T$ as a mass operator giving rise to 
a mass
term. 
Then the massive operator
$
D_M -T : \cH^1(M^{10}, E_0) \to L^2(M^{10}, E_0)
$
is invertible. 

\vspace{3mm}
Note that  a mass term is needed to describe the contribution of the 
Rarita-Schwinger field to the partition function of M-theory on $Y^{11}$
\cite{Flux}. Consider the massive Rarita-Schwinger operator 
$D^m_{RS}=D_{RS}+ im$, where $m$ is a constant so that $im$ is 
a soft perturbation. Different limits are obtained for $m \to \pm \infty$, so that
the determinant is ${\rm det} D_{RS} \exp (\pm i I_{RS})$, with the 
sign depending on the sign of $m$. Here $I_{RS}$ is the Rarita-Schwinger
index in twelve dimensions. Therefore, 
this makes it only natural to work with massive 
Dirac operators in twelve dimensions.

 \vspace{3mm}
 In what follows we aim to characterize the effect of the corner 
 on the phase of the partition function. 
 The matrix $T$
 squares to the identity matrix
 $T^2={\rm Id}$, so that $T$ has eigenvalues $\pm 1$.
Let
  $\Lambda_T \subset {\rm ker}D_M$ be the $+1$-eigenspace of $T$
and 
 let $\Lambda_{\Gamma T}\subset {\rm ker}D_M$ is the $+1$-eigenspace of the 
self-sdjoint unitary automorphism $\Gamma T$. That is,
\(
\Lambda_T:=\{ \Psi \in \ker D_M ~:~ T\Psi = + \Psi  \}\;,
\qquad
\Lambda_{\Gamma T}:=\{ \Psi \in \ker D_M ~:~ \Gamma T\Psi = + \Psi  \}\;.
\)
Now, considering an extension $\widehat{T}$ of $T$ to $Z^{12}$,
the operator 
\(
\widehat{D} -\widehat{T}: x^\alpha \cH^1_b(Z^{12}, E) \to x^\alpha L^2_b(Z^{12}, F)
\)
is Fredholm for all $0 < |\alpha| <\delta$ for some $\delta>0$. 
Recall that $\widehat{Y}_1^{11}$ is formed by attaching an infinite cylinder 
$(-\infty, 0]_{s_2} \times M^{10}$ to the eleven-dimensional compact manifold 
with boundary $Y_1^{11}$, and similarly for  $\widehat{Y}_2^{11}$. 
Let $\widehat{T}_1$ and $\widehat{T}_2$ 
denote the operators induced by $\widehat{T}$ on $Y^{11}_1$ and
$Y^{11}_2$, respectively. 
We need to consider boundary conditions on the spinors at the 
infinite ends of the cylinders. 
The two sets 
\bea
\Lambda_{C_1}=\left\{
\lim_{s_2 \to -\infty} \Psi(s_2, y) : \Psi \in C^\infty (\widehat{Y}^{11}_1, E) 
{\rm ~is~bounded,~and~} \widehat{D}_1 \Psi=0
 \right\}\;,
\\
\Lambda_{C_2}=\left\{
\lim_{s_1 \to -\infty} \Psi(s_1, y) : \Psi \in C^\infty (\widehat{Y}^{11}_2, E) 
{\rm ~is~bounded,~and~} \widehat{D}_2 \Psi=0
 \right\}\;,
\eea
are called the scattering Lagrangian subspaces of $\widehat{D}_1$
and $\widehat{D_2}$, respectively. It turns out that, for each $i=1,2$,
$\Lambda_{C_i} \subset {\rm ker} D_M$ and the 
dimension of $\Lambda_{C_i}$ is exactly one-half the dimension of 
${\rm ker} D_M$ \cite{Mel}.
The scattering matrix of $\widehat{D}_i$ is the operator $C_i: {\rm ker}D_M \to {\rm ker}D_M$
defined by $C_i=+1$ on $\Lambda_{C_i}$ and $C_i=-1$ on $\Lambda_{C_i}^\perp$,
where ``$\perp$" stands for orthogonal 
complement with respect to the $L^2$ inner product. Then $C_1$ and $C_2$ are 
 odd with respect to 
$\Gamma$ (see \cite{Mu2}).

\vspace{3mm}
\paragraph{Effect of the interval and the mass.}
The $b$-eta-invariants and the dimensions of the kernels of the massive operators 
on $Y_i^{11}$ ($i=1,2)$ can be given in terms of their massless counterparts as
\bea
\dim \ker (\widehat{D}_i - \widehat{T}_i)&=& \dim \ker \widehat{D}_i + \dim (\Lambda_{T_i} \cap \Lambda_{C_i})\;,
\\
{}^{\bf b}\eta (\widehat{D}_i - \widehat{T}_i)&=&
{}^{\bf b}\eta (\widehat{D}_i ) \pm \mu (\Lambda_{T_i} , \Lambda_{C_i})\;,
\eea
where $T_1=T$ and $T_2=\Gamma T$, and the upper and lower signs are taken for 
$i=1$ and 2, respectively. 
Here $\mu(\Lambda_T, \Lambda_{C_1})$ and $\mu(\Lambda_{\Gamma T}, \Lambda_{C_2})$
are spectral expressions  \cite{LW}
which can be interpreted as an `exterior angle'
between the Lagrangian  subspaces \cite{Bu}.
Then 
the index in this case is given in terms of the usual (but with $b$-calculus) 
APS terms plus a correction due to the corner, namely 
\cite{LM}
\bea
{\rm ind}_\alpha (\widehat{D} -\widehat{T}) 
&=&\int_{Z^{12}} \widehat{A}(Z^{12})
-\frac{1}{2}\sum_{i=1}^{2} 
\left\{
{}^{\bf b}\eta (\widehat{D}_i- \widehat{T}_i) 
+ {\rm sign}\alpha \cdot {\dim \ker}(\widehat{D}_i  - \widehat{T}_i)
\right\}
\nonumber\\
&=&
\int_{Z^{12}} \widehat{A}(Z^{12})
-\frac{1}{2}\sum_{i=1}^{2} 
\left\{
{}^{\bf b}\eta (\widehat{D}_i) \pm \mu (\Lambda_{T_i} , \Lambda_{C_i})
+ {\rm sign}\alpha \cdot \left( {\dim \ker}(\widehat{D}_i) +\dim (\Lambda_{T_i}\cap \Lambda_{C_i})
\right)
\right\}
\nonumber\\
&=&
\int_{Z^{12}} \widehat{A}(Z^{12})
-\frac{1}{2}\sum_{i=1}^{2} 
\left\{
{}^{\bf b}\eta (\widehat{D}_i) + {\rm sign}\alpha \cdot {\dim \ker}\widehat{D}_i  
\right\}
-\frac{1}{2}c_\alpha (\Lambda_T, \Lambda_{C_1}, \Lambda_{C_2})\;.
\eea
The correction term is given by the expression
\(
c_\alpha (\Lambda_T, \Lambda_{C_1}, \Lambda_{C_2})=
{\rm dim}(\Lambda_T \cap \Lambda_{C_1}) +
\mu (\Lambda_T, \Lambda_{C_1})+
{\rm dim}(\Lambda_{\Gamma T} \cap \Lambda_{C_2})
- \mu(\Lambda_{\Gamma T}, \Lambda_{C_2})\;.
\label{eq corr c}
\)
The precise value of this term will require explicit evaluation for a given situation. 
Generally, we then have

\begin{proposition}
The correction to the phase due to the presence of a corner, for the case when the
Dirac operator on the ten-dimensional corner is not invertible, is given 
$\exp \pi i$ times the term  $c_\alpha (\Lambda_T, \Lambda_{C_1}, \Lambda_{C_2})$.
\end{proposition}


We will see below 
in section \ref{sec sig}
(in terms of the signature) 
 that such a term 
will be zero 
in many simple cases 
of interest.

\subsection{Formulation using the signature operator}
\label{sec sig}

In \cite{DMW-sig} we formulated the phase of the partition function 
by using the signature operator $\cS$ and the signature $\sigma(Z^{12})$ of 
the twelve-manifold $Z^{12}$. That was done for the case when M-theory
is taken on an eleven-manifold without boundary. In this section we will 
consider the extension to the case when there are corners present. 
In section \ref{sig cor}
we consider the case when the relation between the ten-dimensional 
manifold $M^{10}$ and the twelve-dimensional manifold $Z^{12}$ 
is through a `fiber' that is a product of two intervals. Then in section 
\ref{sec rev} we formulate the problem using orientation-reversing 
involutions. The first is related to the formulation of heterotic string theory
as a boundary in M-theory, and the second is related to viewing that 
theory as the base of a circle bundle with an orientation-reversing 
involution on the fiber. This is a global extension of the \cite{HW1} \cite{HW2}
 ``upstairs" and ``downstairs" notions, respectively, to corners.

\subsubsection{The phase of the partition function for manifolds with corners via the signature}
\label{sig cor}

We begin by recalling some properties of the signature that will be useful for us. 
For $Z^{12}$ closed and oriented, Hodge 
theory implies that the index of the signature operator
$\cS$ is given as the integral 
$
 \sigma(Z^{12})={\rm index}(\cS)=\int_{Z^{12}} L$,
where $L$ is the Hirzebruch L-polynomial and the right-hand side 
is 
the signature of the quadratic form on $H^6(Z^{12};\R)$ given by the
cup product \cite{Hir}. 
There is a bilinear form
on $H^{6} (Z^{12}) \otimes H^{6}(Z^{12}) \buildrel{\cup}\over{\longrightarrow}  \R$,
where the cup product $\cup$ is symmetric and nondegenerate, and the signature of $Z^{12}$ is 
$\sigma (Z^{12})= \sigma (\cup)$ with the following properties

\begin{enumerate}
\item {\it Orientation reversal}: $\sigma (-Z^{12})=- \sigma (Z^{12})$.
This will be relevant when we connect to heterotic 
string theory via the Horava-Witten involution \cite{HW1} \cite{HW2}. 

\item {\it Product}: $\sigma (M^{4} \times N^{8})= \sigma(M^{4}) \sigma (N^{8})$.
This will be useful in compactifications to four dimensions and to relating the 
corresponding secondary invariants in eleven dimensions to those in seven dimensions. 

\item {\it Bordism invariance}: If $Z^{12}= \partial W^{13}$ then $\sigma (Z^{12})=0$.
In this case the deficit, that is the $\eta$-invariant, will be given by the L-genus. 

\item {\it Novikov Additivity}: For $Z^{12}= Z_1^{12} 
\bigcup_{\partial Z_1^{12}=\partial Z_2^{12}} Z_2^{12}$, the relation  
$\sigma (Z^{12})= \sigma (Z_1^{12}) + \sigma (Z_2^{12})$ holds 
\cite{No}. We will use a variation on this, that is Wall's non-additivity
\cite{Wa}; see \eqref{eq nono} below.
The common boundary will be (components of) the eleven-manifold on 
which M-theory is studied.

\end{enumerate}

Consider the case when M-theory is take on an eleven-manifold $Y^{11}$
with boundary $\partial Y^{11}=M^{10}$. The partition function in this
case is studied in \cite{DFM}, where also multiple boundaries 
$\partial Y^{11}=\bigcup_i M_i^{10}$ were allowed.  Now take the bounding twelve-manifold 
$Z^{12}$ to be partitioned into two manifolds $Z_1^{12}$
and $Z_2^{12}$, which have boundaries, which in turn have `boundaries',
so that the twelve-manifold $Z^{12}$ is a manifold with corners of 
codimension-two, with the corners being the union of ten-manifolds $M_i^{10}$. 
In this case we use the index theorems for manifolds with corners, 
as constructed in \cite{Bu} \cite{Mu} \cite{HMM}.
What replaces a two-disk $\mathbb{D}^2$ for the fiber over the ten-manifold 
is topologically a square, that is topologically a product of two intervals $I\times I$. 
For simplicity, we will consider unit intervals for the rest of this section.

\vspace{3mm}
We will next consider the corresponding eta-invariants.

\paragraph{Signature eta-invariant for corners.}
Consider $Z^{12}$ to have a boundary $Y^{11}$ which has a neighborhood 
metrically of the form $Y^{11} \times [0,1)$. 
Consider $M^{10}\subset Y^{11}$ as a separating closed ten-dimensional
submanifold which possesses a neighborhood metrically of the form 
$M^{10} \times (-1, 1)$.
In this setting, the corresponding eta-invariant can  be defined as the signature defect
 \cite{Rees} 
\(
\eta(Y^{11}, M^{10}) =\int_{Z^{12}} L_{12} - \sigma (Z^{12})\;.
\)
The ten-manifold 
$M^{10}$ is a corner of $Z^{12}$, that is, $\partial Z^{12}=Y^{11}$ with 
a neighborhood of $Y^{11}-M^{10}$ of the form $(Y^{11}- M^{10})\times [0,1)$
and a neighborhood of $M^{10}$ metrically of the form $M^{10} \times ([0,1) \times [0,1))$.
For two twelve-manifolds $Z_1^{12}$ and $Z_2^{12}$ having $Y^{11}$ as a common 
boundary and $M^{10}$ as a common corner, the invariant $\eta(Y^{11}, M^{10})$
is well-defined \cite{Rees}, that is takes the same value on both twelve-manifolds,
\(
\int_{Z_1^{12}} L_{12} - \sigma (Z_1^{12})= 
 \int_{Z_2^{12}} L_{12} - \sigma (Z_2^{12})\;.
\)
Furthermore, the above $\eta$-invariant of the pair $(Y^{11}, M^{10})$
can be related to the eta-invariant of closed $Y^{11}$ as follows (see \cite{Rees}). 
Let $Y'^{11}=\partial Z^{12}$ be a closed eleven-manifold 
and $M^{10} \in Y'^{11}$ a closed separating submanifold as above.
Assume there exists $Y_0^{11}\subset Z^{12}$, an eleven-dimensional 
 submanifold with boundary $\partial Y_0^{11}= M^{10}=\partial Z^{12} \cap Y_0^{11}$,
 separating $Z^{12}$ into two components $Z_+^{12}$ and $Z_{-}^{12}$ such that the
 neighborhoods of $M^{10}$ in $Z_\pm^{12}$ are metrically of the form 
 $M^{10} \times [0,1) \times [0,1)$ and neighborhoods of $Y_0^{11}$ in 
 $Z_\pm^{12}$ metrically of the form $Y_0^{11} \times [0,1)$. 
  Let $Y_\pm^{11}$ be the component of $Y^{11}-M^{10}$ lying in $Z_\pm^{12}$,
 so that $(Y_\pm^{11} \cup_M Y_0^{11}=N_{\pm}^{11}, M^{10})$ defines 
 a pair of manifolds as above for each of $\pm$. Then the eta-invariants are
 related as follows
 \(
 \eta (Y^{11})= \eta(Y_+^{11}, M^{10}) -
 \eta (Y_{-}^{11}, M^{10}) + \delta (M^{10}; Y_{-}^{11}, Y_0^{11}, Y_+^{11})\;,
\label{clop}
 \)
where $\delta$ is Wall's invariant, that is the obstruction to additivity of the signature
\cite{Wa}
\(
\sigma (Z_+^{12} \cup_{Y_0^{11}} (-Z_{-}^{12}))=
\sigma (Z_{-}^{12}) - \sigma (Z_+^{12})
+
\delta (M^{10}; Y_{-}^{11}, Y_0^{11}, Y_+^{11})\;.
\label{eq nono}
\)
Therefore, we see that we can `trade' the boundary in M-theory with Wall's 
invariant. However, this invariant involves the homology
groups $H_5(M^{10})$ and $H_5(Y^{11})$, or dually 
the cohomology groups
$H^5(M^{10})$ and $H^6(Y^{11})$; since M-theory does not support
fields in these degrees, this results in the Wall invariant being zero in this
case. Then the eta-invariant of the eleven-manifolds with boundary can 
be expressed in terms of the eta-invariants of the closed eleven-manifold, 
that is, \eqref{clop} reduces to 
\(
 \eta (Y^{11})= \eta(Y_+^{11}, M^{10}) -
 \eta (Y_{-}^{11}, M^{10})\;.
\)
Note that the phase factor of the partition function in the presence of a boundary 
has the same expression as when there is no boundary, except of course that
the terms would be a modification for the meaning of the terms \cite{DFM} \cite{FM}. 
Therefore, the above result might not be surprising.

\begin{proposition}
The correction term to the phase in the signature formulation is given by Wall's
non-additivity term. When $H^5(M^{10})$ and $H^6(Y^{11})$ are zero, then there
is no correction. 
\end{proposition}

\paragraph{Multiple boundary components.}
Let $Y_i^{11}$, $i=1, \cdots, N$ be an ordering of codimension-one
boundary components of $Z^{12}$. The intersections $M_{ij}^{10}:= Y_i^{11}\cap Y_j^{11}$,
which need not be connected,  
are the codimension-two boundaries, i.e. the corners. 
Let $D_{ij}$ denote the Dirac operator on $M_{ij}^{10}$ induced by either 
$D_i$ on $Y^{11}_i$ or $D_j$ on $Y^{11}_j$. The two operators are
 the same up to sign and
are induced by the signature operator on $Z^{12}$. 
Note that the signature itself can be 
viewed as a generalized Dirac operator. 
Assuming the APS boundary conditions, the signature can be written as \cite{HMM}
\(
\sigma (Z^{12})= \int_{Z^{12}} L_{12} -\frac{1}{2}\left( 
\sum_{i=1}^N \eta (D_i) + \frac{1}{i\pi} {\rm tr} P_\Lambda
\right)\;,
\label{eq HMM}
\)
where $P_\Lambda$ is the corner analytical correction term involving a
certain projection matrix $P_\Lambda$ (see \cite{HMM} for details). 
This is, in a sense, an analog of the correction term 
\eqref{eq corr c} in the Dirac operator case.
We will not use the explicit form of this correction 
term in this paper, and below we consider
 examples where this is actually zero.
Note that the number of zero modes does not appear in \eqref{eq HMM} since
they are absorbed in the index, since this is defined with respect to 
a modified projection, called the augmented  APS projection \cite{HMM}
(see also the remark right after proposition \ref{prop etabar}).

\paragraph{Example 5. Products.}
Consider $Z^{12}$ as the product of two manifolds with boundary, so that 
$Z^{12}$ is a manifold with corners of codimension-two.
In this case $N=2$. 
Let $X^4$ and $U^8$ be two manifolds with product $b$-metrics 
${}^{\bf b}g_X$ and ${}^{\bf b}g_U$ and boundaries $\partial X^4=Y^3$ and 
$\partial U^8= V^7$, respectively.  
The signatures of $X^4$ and $U^8$ are given by 
\(
\sigma (X^4)=\int_{X^4} L_4 -\frac{1}{2}\eta (Y^3)\;,
\qquad 
\sigma (U^8)=\int_{U^8} L_8 -\frac{1}{2}\eta (V^7)\;.
\)
The signature of $X^4 \times U^8$ is the product of 
signatures of the factors
\bea
\sigma (X^4 \times U^8) &=&
\int_{X^4 \times U^8} L_3 -\frac{1}{2} \eta (Y^3) \int_{U^8}L_8 
-
\frac{1}{2}\eta (V^7) \int_{X^4} L_2 
+
\frac{1}{4} \eta (Y^3) \eta (V^7)
\nonumber\\
&=&
\int_{X^4 \times U^8} L_{12} -\frac{1}{2} \eta (Y^3) \sigma (U^8)
-
\frac{1}{2}\eta (V^7) \sigma(X^4)
-
\frac{1}{4} \eta (Y^3) \eta (V^7)\;.
\eea
The $b$-eta-invariants are given by 
\(
{}^{\bf b}\eta(X^4 \times V^7) + {}^{\bf b}\eta(Y^3 \times U^8) =
\sigma (X^4)~ {}^{\bf b}\eta (V^7) + 
\sigma (U^8)~ {}^{\bf b}\eta (Y^3) +
\eta (Y^3) \eta (V^7)
\)
and account for the eta-terms in \eqref{eq HMM}.
The corner is $M^{10}=Y^3 \times V^7= \partial (X^4 \times V^7) - \partial (Y^3 \times U^8)$.
Identify the Lagrangians $\Lambda_{12}$ and $\Lambda_{21}$, which are the 
asymptotic limit of solutions of the generalized Dirac equation
$D \psi=0$ on $X^4 \times V^7$ and 
$Y^3 \times U^8$, respectively. 
Using $K_1 =\ker (D_Y)$, $K_2=\ker (D_V)$ and 
$K = K_1 \otimes K_2 =\ker (D_{Y \times V})$, and 
$\Lambda_i \subset K_i$ denote the scattering Lagrangian for 
$D_Y$ and $D_V$, respectively. A calculation using heat kernels 
shows that the corner correction term vanishes in this case 
(see \cite{HMM}).

\paragraph{Example 6. Disks.} Let us consider a very simple, but physically realistic case,
where $X^4=\mathbb{D}^4$ and $U^8=\mathbb{D}^8$, the 4- and 8-dimensional 
disks, respectively. Assume then that the corresponding boundaries
$Y^3=S^3$ and $V^7=S^7$, the round 3- and 7-spheres. Then in this case
the eta-invariants vanish and so the signature formula for manifolds with 
corners reduces drastically all the way to that of a manifold without boundary, that is 
$\sigma (\mathbb{D}^4 \times \mathbb{D}^8) =\int_{\mathbb{D}^4}L_1 \int_{\mathbb{D}^8}L_2$. 
But then these are given in terms of Pontrjagin 
 classes, which are zero in cohomology for disks.

\begin{proposition}
The phase of the partition function in the case of a product 
$Z^{12}= X^4 \times U^8$ of two manifolds with boundary $X^4$ and 
$U^8$ is given by the Atiyah-Patodi-Singer index; that is, there is no corner 
correction in this case. 
\end{proposition}

\subsubsection{Orientation-reversing involutions}
\label{sec rev}

In this section we consider the ``downstairs" formulation of the heterotic theory, that is 
the case when $Y^{11}=M^{10}\times S^1/\Z_2$ with an orientation-reversing 
involution $\pi$ on $S^1$ generated by $\Z_2$. We will in fact work in more generality
in what follows.

\vspace{3mm}
Let $Z^{12}$ be a compact oriented smooth twelve-manifold 
with boundary, $Y^{11}$. 
This boundary is called a {\it reflecting boundary} of $Z^{12}$ if it admits an orientation-reversing
involution $\pi$. A simple example of a reflecting boundary of $Z^{12}$ is 
an eleven-sphere $S^{11}$. 
The {\it doubling} of the manifold $Z^{12}$ with a reflecting boundary $(Y^{11}, \pi)$
is a $C^\infty$-homeomorphism 
$h: Z^{12} \to N^{12}$ 
where $N^{12}$ is a smooth 
closed manifold with an involution $\nu : N^{12} \to N^{12}$ such that 
we have an equality of compositions 
$
\nu\circ h \circ \iota= h \circ \iota \circ \pi : Y^{11} \to N^{12}
$
in the diagram 
\(
\xymatrix{
N^{12} 
\ar[d]^\nu
&&
Z^{12}
\ar[d]^{=}
\ar[ll]_h
&&
~~Y^{11}
 \ar@{_{(}->}[ll]_\iota
\ar[d]^\pi
\\
N^{12} 
&&
Z^{12}
\ar[ll]_h
&&
~~Y^{11}
 \ar@{_{(}->}[ll]_\iota
}\;.
\)
A {\it symmetric} metric on 
the double $N^{12}$ is a Riemannian metric for which the involution 
$\nu$ is an isometry. Starting from a smooth Riemannian metric $g$ on 
$N^{12}$ one gets a smooth symmetric metric  by setting, for $x \in N^{12}$, 
(see \cite{Hs})
\(
g(x)= \frac{1}{2} \left[ 
g(x) + g(\nu (x))
\right]\;.
\)
Let the curvature of $g$ be $R_g$ and consider  the 
corresponding Pontrjagin forms 
$p_i(g)$, $i=1, 2,3$.
The Hirzebruch signature theorem then gives
\(
\sigma (N^{12})= \int_{N^{12}} L_{12}(p_1 (g), p_2(g), p_3(g))\;. 
\)
The symmetric metric $g$ on $N^{12}$ restricts to a metric $g_Z$ on 
$Z^{12}$, also called 
a symmetric metric. Then, applying \cite{Hs}, the signature
of $Z^{12}$ with a reflecting boundary $Y^{11}$
is given by 
\(
\sigma (Z^{12}, Y^{11})=\int_{Z^{12}} 
L_{12}(p_1 ({g_Z}), p_2({g_Z}), p_3({g_Z}))\;.
\)
This reduces to the Hirzebruch signature theorem when $Y^{11}$ is empty.
Therefore, in the presence of a reflecting boundary, the
Chern-Simons and one-loop terms are encoded in primary characteristic classes
and the eta-invariant drops out. 

\begin{proposition}
The signature part of the phase of the partition function is 1 when 
$Y^{11}$ has a reflecting boundary. 
\end{proposition}

It would be interesting to work out examples where the correction terms
are evaluated explicitly. 
We also plan to extend the discussion in this paper to more refined invariants.

\vspace{0.5cm}
\noindent {\bf \large Acknowledgements}

\vspace{2mm}
\noindent The author would like to thank
Sergei Novikov and Jonathan Rosenberg for useful 
discussions on the signature, and Andrew Hassell for useful remarks
on analysis on manifolds with corners. 
The author also thanks 
the Department of Mathematics 
at the
 University of 
Melbourne and both the Department of Mathematics and the Department 
of Theoretical Physics at the Australian National University 
for their hospitality 
during the writing of 
 this paper.


\end{document}